\documentstyle[epsfig]{aipproc}

\begin{document}
\title{Active-Sterile Neutrino Transformation\\ and r-Process Nucleosynthesis}

\author{G. C. McLaughlin}
\address{TRIUMF, 4004 Wesbrook Mall, Vancouver, BC, V6T2A3}

\maketitle

\begin{abstract}
The type II supernova is considered as a candidate site for the
production of heavy elements. Since the supernova produces an 
intense neutrino flux, neutrino scattering processes will impact 
element formation.  We examine active-sterile neutrino conversion 
in this  environment and find that it  may 
help to produce the requisite neutron-to-seed ratio for 
synthesis of the r-process elements.  
  
\end{abstract}


The r-process of nucleosynthesis accounts for the most neutron rich of the 
heavy elements.  The most likely environment for this type of synthesis
is the late time ($t > 10 \, {\rm s}$ post-core bounce) supernova environment.
Many studies have explored this \lq neutrino driven wind\rq\ as a 
candidate environment and found it to be potentially viable \cite{r1,me92}.  
However, to date, no model correctly reproduces
the observed abundance pattern.

In the neutrino driven wind, material in the form of free nucleons is
\lq lifted\rq\ off of the surface of the neutron star by energy deposited
by neutrino interactions.  Analytic and semianalytic parameterizations
of the thermodynamic and hydrodynamic conditions in the wind can be
obtained \cite{dun,qw}.  Models of this type may be used to explore the 
range of conditions within the context of the wind which will produce 
the solar system distribution of r-process elements. The key determinant
of whether a given scenario will produce the r-process is the 
neutron to seed nucleus ratio at the onset of the neutron capture phase.  
This ratio must be quite high ($ R > 100$) in order to produce 
the very neutron-rich r-process elements.  The factors which determine
the neutron-to-seed ratio are the entropy of the material, the
hydrodynamic outflow timescale and the electron fraction, 
$Y_e = 1/(1 + n/p)$ where $n/p$ is the neutron-to-proton
ratio.  A
study of many possible model parameters shows that one must decrease the
electron fraction, and/or increase the entropy 
and/or decrease the hydrodynamic outflow timescale,
relative to the conditions found in typical wind models, in
order to produce the neutron-to-seed ratio necessary for the 
r-process \cite{me97,hwq}.  

Including the effects of neutrino interactions in general tends to make
the requisite conditions for r-process element production more extreme
\cite{me95,fm95}.  In particular, neutrino capture on free
nucleons during alpha particle formation increases the electron
fraction \cite{fm95}.  This is the ``alpha effect''.  Other neutrino 
process are discussed in \cite{me92,wick}.  

There are three possible solutions to this problem.  The first is that
the supernova is the site of r-process synthesis, but it does not
occur in the neutrino driven wind as it is currently modeled.  The
second is that the r-process elements are made at some other site such as
neutron star-neutron star mergers.  However, timescale arguments
combined with isotopic abundance measurements and observations
of old halo star metallicity   
show that this site is unlikely to account for the entire r-process 
distribution \cite{wasserburg,sneden}.
  
The third solution is the one that is investigated here:
active-sterile ($\nu_e \leftrightarrow \nu_s$, $\bar{\nu_e}
\leftrightarrow \bar{\nu_s}$) neutrino transformation.  The $\nu_s$ in
our study is defined as a particle which mixes with the $\nu_e$ (and
possibly also with $\nu_\mu$, and/or $\nu_\tau$) but does not contribute to
the width of the Z boson. 

If we neglect $\nu$-$\nu$ forward scattering contributions to the
weak potentials, then the equation which governs the evolution of the 
neutrinos as they pass though the material in the wind can be written as:

\begin{equation}
i\hbar \frac{\partial}{\partial r} \left[\begin{array}{cc} \Psi_e(r)
\\ \\ \Psi_s(r) \end{array}\right] = \left[\begin{array}{cc}
\varphi_e(r) & (\delta m^2 / 4 E)\sin{2\theta_v}
 \\ \\ (\delta m^2 / 4 E)\sin{2\theta_v} & -\varphi_e(r)
\end{array}\right]
\left[\begin{array}{cc} \Psi_e(r) \\ \\ \Psi_s(r)
  \end{array}\right]\,,
\label{eq:msw}
\end{equation}
where
\begin{equation}
  \label{2} \varphi_e(r) = \frac{1}{4 E} \left( \pm
 2 \sqrt{2}\ G_F \left[
  N_e^-(r) - N_e^+(r) - \frac{N_n(r)}{2} \right] E - \delta m^2
  \cos{2\theta_v} \right)
\label{eq:potne}
\end{equation}
The upper sign is relevant for neutrino transformations; the lower one is
for antineutrinos.  In these equations
$\delta m^2 \equiv m_2^2 - m_1^2$ is the vacuum mass-squared
splitting, $\theta_v$ is the vacuum mixing angle, $G_F$ is the Fermi
constant, and $N_e^-(r)$, $N_e^+(r)$, and $N_n(r)$ are the 
total proper number densities of electrons, positrons, 
and neutrons respectively in the
medium.   Resonances can occur when the on-diagonal terms in the wave equation
are zero.  The quantity in the brackets in Eq. \ref{eq:potne} is proportional
to $Y_e - 1/3$.
Since the electron fraction can take on values between zero and one,
the bracketed quantity in  Eq. \ref{eq:potne}  can be either positive or negative.  
Therefore, for a given choice of $\delta m^2$, 
 resonances may occur for neutrinos or antineutrinos depending
on the value of the electron fraction. 

In order to determine the survival probabilities of the neutrinos
and antineutrinos, we must know the electron fraction.
In the neutrino driven wind, neutrino and antineutrino capture
are the most important reactions in determining the electron
fraction. However, near the surface of the protoneutron star there is
also a contribution from electrons and positrons:  
\begin{equation}
\label{eq:ccf}
\nu_e + {\rm n}  \rightleftharpoons {\rm p}+ e^{-}; \; \; \;
\bar{\nu}_e + {\rm p} \rightleftharpoons {\rm n} + e^{+}.
\end{equation}
Therefore, the problem involves a feedback mechanism.  The neutrino
capture rates determine the electron fraction and the electron fraction
determines the potential which is used in the neutrino transformation
 equations.  These
equations then determine survival probabilities of the neutrinos, and
therefore their capture rates.

We note that we can not assume weak equilibrium, 
(very fast capture rates compared with outflow rate) or weak freeze-out 
(very slow rates compared with the outflow rate).  The rates must be tracked
numerically.  A previous study took the limit of weak equilibrium 
 and found different behavior \cite{nunokawa} than presented here.

We perform  calculations by tracking a mass element in the
neutrino driven wind.  We use the results of analytic models
\cite{qw}, where $r \propto \exp(-t/\tau)$ and $\rho
\propto r^{-3}$ where $\tau$ is the outflow timescale. Close
to the surface, before the wind begins to operate we use the density
profile of Wilson and Mayle \cite{wilson}.  At each time step we calculate 
all thermodynamic quantities, calculate the
weak rates and evolve the neutrino transformation  
equations forward.  We assume,
 since the outflow timescale is short, $t \sim \tau \lesssim 0.5 \, {\rm s}$,  
that each mass element will have the same evolution as the
previous one.  Since we use
a nuclear statistical equilibrium calculation, we cut our calculations
off at the time when heavy nuclei begin to form.  More detail is contained
in \cite{us}.

\begin{figure}[b!] 
\centerline{\epsfig{file=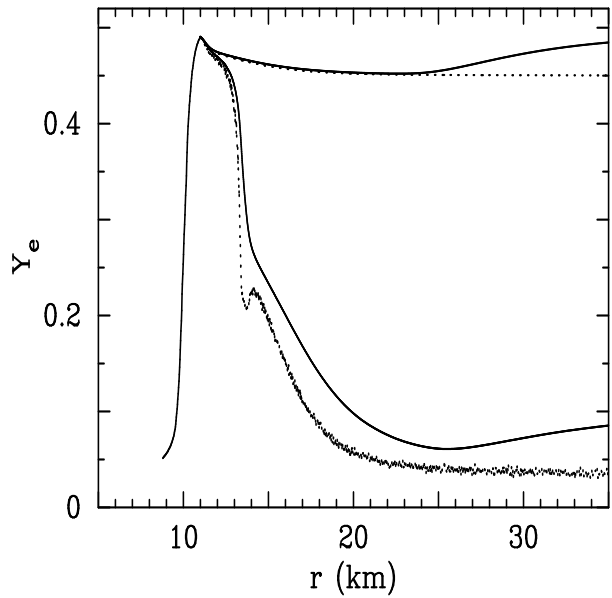,height=3.5in,width=3.5in,angle=-90}}
\vspace{10pt}
\caption{The electron fraction is plotted against distance from the
center of the neutron star.  The upper line shows the evolution with no 
transformation.  The lower line shows the evolution of active-sterile mixing
parameters of $\sin^2 \theta = 0.01$ and $\delta m^2 = 20 \, {\rm eV}^2$.
The dotted line shows the value that the electron fraction would take
on if weak equilibrium obtained. 
\label{fig:efraction1}}
\end{figure}

A calculation for one mass element is shown in Fig. \ref{fig:efraction1}.  The
upper curve shows the evolution for the case of no neutrino oscillations.
The wind parameters were $ \tau = 0.3 \, {\rm s}$, $s = 100$.  The solid line shows 
the actual electron fraction, while the dotted line shows what the
electron fraction would be if weak equilibrium obtained.  The initial rise is
due to Pauli unblocking of the electrons at the surface of the proto-neutron
star.   The lower curve
shows the evolution of the electron fraction for mixing parameters of 
$\delta m^2 = 20 \, {\rm eV}^2$, $\sin^2 2 \theta_{\rm v} = 0.01$.  In the latter
case there is a rapid drop in the electron fraction when the neutrinos begin to 
transform.  In fact, there are three neutrino transformations.  Initially, 
the electron 
neutrinos transform to steriles.  Later antineutrinos transform,
and finally, when the density falls far enough the antineutrinos transform back.
The first transformation of the antineutrinos is seen in the
small bump in the equilibrium electron fraction.  A small ``alpha effect'' can
be see as the slight rise in both solid lines at large distance.

\begin{figure}[b!] 
\centerline{\epsfig{file=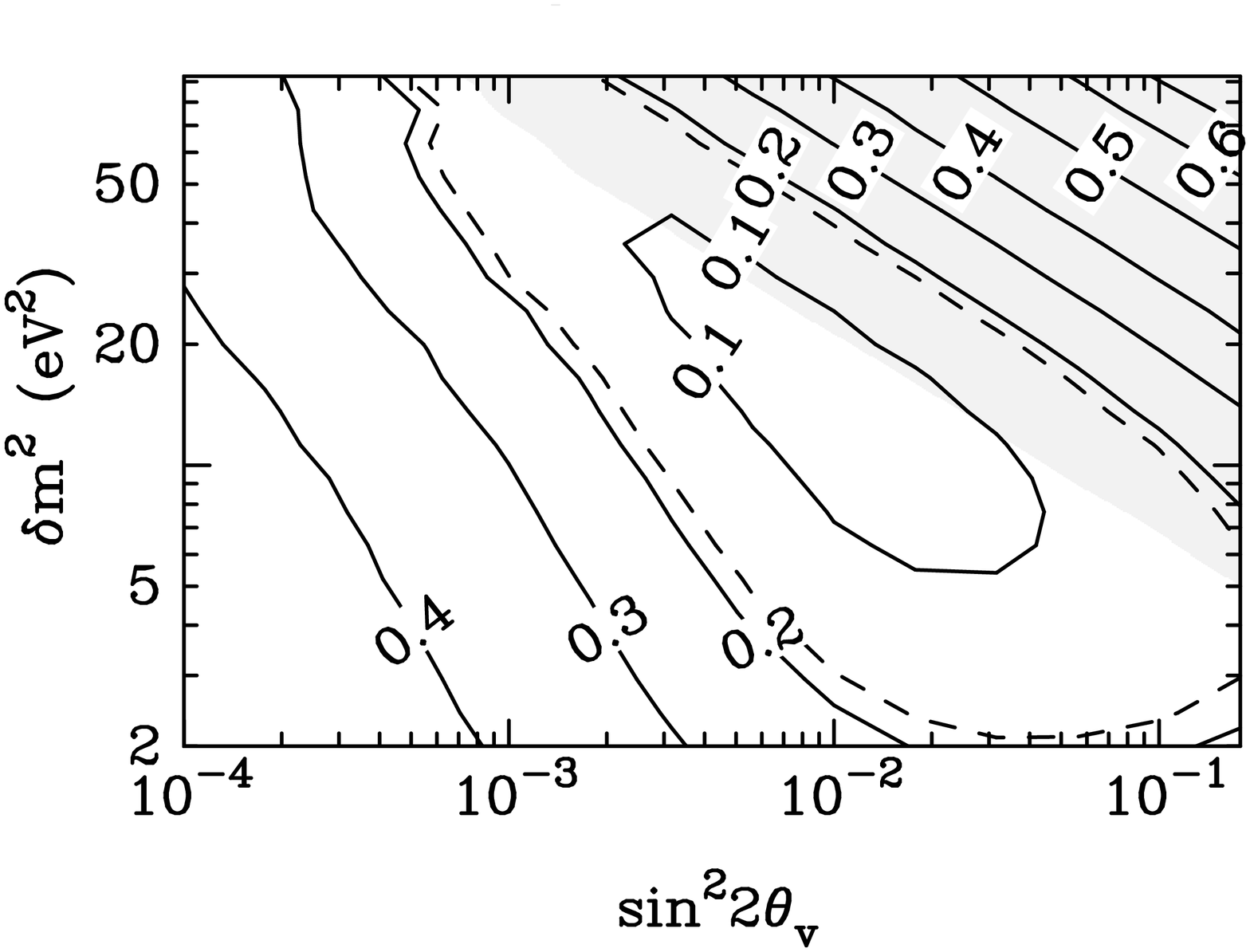,height=3.5in,width=3.5in,angle=-90}}
\vspace{10pt}
\caption{Contour plot of electron fraction as measured at the
point where heavy nuclei begin to form.  Neutrino driven wind parameters
employed here are $s/k = 100$, $\tau = 0.3 {\rm s}$. 
\label{fig:efraction2}.}
\end{figure}

The active-sterile transformation scenario successfully reproduces a low electron
fraction which is beneficial to the r-process.  We now consider a
range of $\delta m^2$, $\sin^2 \theta$ parameters. A contour plot of the
electron fraction is shown in Fig. \ref{fig:efraction2}.  Inside the
dashed contour shows the region where conditions are neutron-rich enough to
be favorable for r-process nucleosynthesis. In the bottom left 
corner of the plot, the solution is approaching the
case without neutrino transformation.  


Although not shown here, we have studied a range of timescales 
for the neutrino driven wind models,
and seen that the qualitative features of this effect
are reproduced \cite{us}. 
 
We have used several approximations in our calculations, which
we are continuing to study.  These include the importance
of the neutrino-neutrino scattering background in the oscillation
equations, the importance of nonradial paths and feedback in the 
dense region near the proto-neutron star.  The region where the latter
two are important is the shaded region in Fig. \ref{fig:efraction2}.
These problems are being studied in \cite{us2}.

{\bf Conclusions:} 
Meteoritic and observational evidence points to supernovae as
the source of the r-process elements, although a 
self-consistent model of the neutrino driven wind which will produce the
r-process elements is still elusive.
In the next few years significant advances are expected 
in supernova modeling.  If and when potential 
hydrodynamic solutions are exhausted
and the caveats above have been explored, then the
 r-process may provide a 
signature for new neutrino physics.

\section*{ACKNOWLEDGEMENTS}

This work was done in collaboration with J. M. Fetter, A. B. Balantekin and
G. M. Fuller.

\end{document}